\documentclass[pdftex]{pasj01} 
\usepackage[dvipdfmx]{}
\usepackage{pslatex} 
\usepackage{color}

\Received{2019/12/24}
\Accepted{2020/5/2}
 
\usepackage{color}
\usepackage{lscape}
\newcommand{\twelveco}{\mbox{$^{12}$CO}} 
\newcommand{\thirteenco}{\mbox{$^{13}$CO}} 
\newcommand{\twelvecoh}{\mbox{$^{12}$CO($J$=2--1)}} 
 
\newcommand{\thirteencoh}{\mbox{$^{13}$CO($J$=2--1)}} 
\newcommand{\twelvecol}{\mbox{$^{12}$CO($J$=1--0)}} 
\newcommand{\thirteencol}{\mbox{$^{13}$CO($J$=1--0)}}

\newcommand {\msun}{\mbox{$M_\odot$}}
\newcommand {\kms}{\mbox{km~s$^{-1}$}}
\newcommand {\kkms}{\mbox{K~km~s$^{-1}$}}
\newcommand {\vlsr}{\mbox{$V_{LSR}$}}
\newcommand {\nhtwo}{\mbox{$N_\mathrm{H_2}$}}
\newcommand {\nco}{\mbox{$N_\mathrm{^{13}CO}$}}
\newcommand {\tmb}{\mbox{$T_\mathrm{mb}$}}

\newcommand {\vcen}{\mbox{$V_\mathrm{center}$}}
\newcommand {\tex}{\mbox{$T_\mathrm{ex}$}}
\newcommand {\tbg}{\mbox{$T_\mathrm{bg}$}}
\newcommand {\kb}{\mbox{$k_\mathrm{B}$}}
\newcommand {\mum}{\mbox{$\mu_\mathrm{m}$}}
\newcommand {\mpro}{\mbox{$m_\mathrm{p}$}}
\newcommand {\dv}{\mbox{$dV$}}

\newcommand {\htwo}{\mbox{H{\sc ii}}}

\begin{document} 

\title{ 
High-mass star formation in Orion B triggered by cloud-cloud collision: Merging molecular clouds in NGC~2024}

\author{Rei \textsc{Enokiya}\altaffilmark{1}, Akio \textsc{Ohama}\altaffilmark{1}, Rin \textsc{Yamada}\altaffilmark{1}, Hidetoshi \textsc{Sano}\altaffilmark{1}, Shinji \textsc{Fujita}\altaffilmark{1}, Katsuhiro \textsc{Hayashi}\altaffilmark{1}, Daichi \textsc{Tsutsumi}\altaffilmark{1},  Kazufumi \textsc{Torii}\altaffilmark{2}, Atsushi \textsc{Nishimura}\altaffilmark{3}, Ryotaro \textsc{Konishi}\altaffilmark{3}, Hiroaki \textsc{Yamamoto}\altaffilmark{1}, Kengo \textsc{Tachihara}\altaffilmark{1} Yutaka \textsc{Hasagawa}\altaffilmark{4}, Kimihiro \textsc{Kimura}\altaffilmark{1, 3}, Hideo \textsc{Ogawa}\altaffilmark{3} and Yasuo \textsc{Fukui}\altaffilmark{1}}
\altaffiltext{1}{Department of Physics, Nagoya University, Furo-cho Chikusa-ku Nagoya, 464-8602, Japan}
\altaffiltext{2}{Nobeyama Radio Observatory, 462-2 Nobeyama, Minamimaki Minamisaku, Nagano, 384-1305, Japan}
\altaffiltext{3}{Department of Physical Science, Graduate School of Science, Osaka Prefecture University}
\altaffiltext{4}{Institute of Space and Astronautical Science, Japan Aerospace Exploration Agency(JAXA)(ISAS)}
\email{enokiya@a.phys.nagoya-u.ac.jp}


\maketitle

\begin{abstract}
 We performed new comprehensive \thirteencoh ~observations toward NGC~2024, the most active star forming region in Orion B, with an angular resolution of $\sim$100\arcsec ~obtained with NANTEN2. We found that the associated cloud consists of two independent velocity components. The components are physically connected to the H{\sc ii} region as evidenced by their close correlation with the dark lanes and the emission nebulosity. The two components show complementary distribution with a displacement of $\sim$0.6 pc. Such complementary distribution is typical to colliding clouds discovered in regions of high-mass star formation. We hypothesize that a cloud-cloud collision between the two components triggered the formation of the late O-type stars and early B stars localized within 0.3 pc of the cloud peak. The duration time of the collision is estimated to be 0.3 million years from a ratio of the displacement and the relative velocity $\sim$3 \kms ~corrected for probable projection. The high column density of the colliding cloud $\sim$10$^{23}$ cm$^{-2}$ is similar to those in the other high-mass star clusters in RCW 38, Westerlund 2, NGC 3603, and M42, which are likely formed under trigger by cloud-cloud collision. The present results provide an additional piece of evidence favorable to high-mass star formation by a major cloud-cloud collision in Orion.
\end{abstract}

\section{Introduction}\label{intro}
\subsection{The Orion region and high-mass star formation}
The Orion region is the nearest most outstanding high-mass star forming region in the solar neighborhood. The H{\sc ii} region NGC~2024 located at 410 pc is the second active H{\sc ii} region next to M42 in Orion and is associated with a reflection nebula NGC~2023 (e.g., \cite{ant82,men07}). 
NGC~2024 is therefore one of the most interesting regions in studying high-mass star formation. A review of the Orion B region is given by \citet{mey08} and references therein.

Observations in radio continuum radiation and recombination lines \citep{kru82, bar89,bik03,ron03} as well as infrared observations indicate that NGC~2024 is ionized by an exciting stars of O8V-B2V \citep{lad91,com96,gia00,hai00,bik03,kan07}.
In addition to the high-mass star, hundreds of low-mass stars are identified in the cluster by near-infrared and X ray observations \citep{lad91,ski03}. These cluster members consist of at least 300 stars including late O-type stars and early B stars. Most of them are suggested to be in mass accretion phase by 10 $\mu$m observations \citep{hai00,hai01}. Although large visual extinction of 30 magnitudes toward the region (e.g., \cite{joh06}) hampers firmly identifying candidates for the ionizing star of NGC~2024, \citet{bik03} identified the exciting star named IRS2b by infrared photometry.

 It is probable that the cluster is smaller than the Orion Nebula Cluster which contains $\sim$10 O- / early B-type stars as well as $\sim$2000 member stars \citep{kro01,ban18}. Near infrared observations of NGC~2024 was used to construct an HR diagram of these stars and the age is estimated to be 1 Myrs or less \citep{get14}. Far infrared observations revealed dense protostellar condensations with a compact distribution, suggesting some external triggering to form them while details remain elusive due to high extinction (e.g., \cite{meg16}.

In order to explain the formation of the OB associations in the Orion region, \citet{elm77} presented a scenario of sequential star formation. In the scenario, an ionization shock front driven by OB stars compresses molecular gas in a direction where molecular gas is distributed, often elongated along the Galactic plane, and after a passage of $\sim$20 pc a compressed layer forms OB stars due to gravitational instability. This scenario explains the age sequence of subgroups of OB associations; in case of Orion the four subgroups Orion Ia, Ib, Ic and Id are separated by 20 pc in the order of age. \citet{bro94} made an extensive study of stars in the Orion region and derived new ages and IMFs of the Orion OB associations. Their results showed that the age of Ic is younger than Ib, contrary to the previous results by \citet{bla65}, which showed the opposite age sequence.

The work raised that the conventional sequential scenario requires reconsideration or modification, even if the scenario may still be largely applicable. On a smaller scale of individual star formation \citet{lee09} presented a study which suggests that Ori-Eri superbubble may be triggering star formation in the apparently interacting small clouds in part of the Orion clouds. We do not have yet a full picture of OB star formation in Orion which assembled gas and stellar datasets comprehensively. So, the formation of the OB association is not fully understood and remains as an open issue.

In the meantime a new picture of star formation in magnetically driven filaments were proposed by \citet{per12}. Most recently, \citet{fuk18b} made a detailed analysis of the \twelvecol ~data in the Orion A cloud taken with the NRO 45 m telescope by \citet{shi11}, and presented an analysis that the Orion A cloud comprises two spatially overlapping components of different velocities. 
These authors suggested that the ionizing stars of M42 and M43 in Orion A were formed by triggering in collision between two clouds of 7 \kms ~velocity difference based on that the two clouds show complementary distribution with a systematic spatial displacement, a typical signature of colliding clouds, for M43.
In the Orion B cloud, in NGC~2024, the JCMT covers in CO ($J$=3--2) only $10.8 \times 22.5$ arcmin$^2$ \citep{buc10} and a large-scale view of the molecular gas has not been revealed. 

\subsection{The orion B cloud}
The Orion B cloud also known as the L1630 cloud was not a subject of many large-scale molecular observations. 
\citet{bal91} observed the cloud in the \thirteencol ~transition and \citet{aoy01} in the C$^{18}$O $J$=1--0 and HCO$^+$ $J$=1--0 transitions.
\citet{kra96} performed \twelveco ~and \thirteenco ~$J$=2--1 and 3--2 observations of the southern part of the Orion B region covering a $\sim$40\arcmin $\times$ 70\arcmin ~area. They revealed physical conditions of molecular clouds in the region for the first time, however the angular resolution and the observing grid is not high (125\arcsec ~and 2\arcmin, respectively) and thus the distribution was coarse and contrasts of cloud intensities were unclear.

\citet{rip13}  used the FCRAO 14 m telescope to map the cloud in the \twelvecol ~transition at higher resolution. 
The \twelvecoh ~transition was observed at 9\arcmin ~resolution in a large scale by \citet{sak94} and \citet{wil05} and was used to derive density and temperature through a comparison with the \twelvecol ~transition. 
\citet{lad91b} preformed an unbiased, systematic survey for dense cores within the Orion B molecular cloud by CS $J$=2--1 transition,  and \citet{ike09} carried out an H\thirteenco$^{+}$ $J$=1--0 core survey in a large area of $1$ deg$^2$.
\citet{mie94} presented an investigation of the statistical properties of fluctuating gas motions in Orion B.
Recently, \citet{nis15}  made a higher resolution comparative study of the \twelvecol ~and $J$=2--1 transitions by using the data taken with the OPU 1.85 m and NANTEN, and derived density and temperature distributions over the whole Orion region including Orion A and Orion B at 3\arcmin ~resolution. 

In order to reveal detailed gas kinematics in NGC~2024, we carried out new observations in the \twelveco ~and \thirteenco $J$=2--1 transitions at 1.5\arcmin ~and 0.079 \kms ~angular and velocity resolutions over a large area including whole the NGC~2024 cloud. As \citet{fuk18b} suggested, the whole Orion region has two velocity components and velocities of these two components are very close (typically less than a few \kms). 
Thus, at least a velocity resolution better than 1 \kms ~is required to resolve them. None of literatures have investigated cloud dynamics in NGC2024 with such finer velocity resolution.
Section 2 gives details of the observations, Section 3 describes the observational results, Section 4 presents discussion on cloud-cloud collision, and Section 5 concludes the paper.

\section{Observation}
Observations of the \thirteencoh ~transition were made with NANTEN2 over an area of 0\fdg5 $\times$ 0\fdg5 in $l$ and $b$. 
The data were taken in a period from December 11, 2016 to December 15, 2016. 
The transition was observed simultaneously with the \twelvecoh ~emission in the on-the-fly mode in 0.8 second integration per a point with a 30\arcsec ~grid spacing. 
The system noise temperature in the Double Side Band was 160 - 220 K toward the zenith. The backend was a digital spectrometer having a band width and resolution of 1 GHz and 61 kHz, respectively. 
These correspond to a velocity coverage of 1300 \kms ~and a velocity resolution of 0.079 \kms.
After convolution with a 2-dimensional Gaussian kernel of 54$\arcsec$, the final beam size was $\sim$ 105$\arcsec$ (FWHM).
Pointing accuracy was measured toward IRC 10216  [$\alpha_\mathrm{J2000}$ = $9^{\mathrm{h}}~47^{\mathrm{m}}~57\fs406$, $\delta_\mathrm{J2000}$ = $-13{^\circ}~16\arcmin~43\fs56$] everyday and confirmed to be better than 10\arcsec. The absolute intensity scale was established by observing OriKL  [$\alpha_\mathrm{J2000}$ = $5^{\mathrm{h}}~35^{\mathrm{m}}~13\fs5$, $\delta_\mathrm{J2000}$ = $-5{^\circ}~22\arcmin~27\fs6$] every hour. 
The final rms noise level was 0.50 K~ch$^{-1}$ at a velocity resolution of 0.079 \kms. We use the Galactic coordinate in the present paper.

\section{Results}
In this section, we present gas distribution toward NGC~2024. We first review global distribution and properties of molecular gas toward this region.
\subsection{Global gas distribution in the NGC~2023/ NGC~2024 region}
\begin{figure*}
\begin{center}
 \includegraphics[width=14cm]{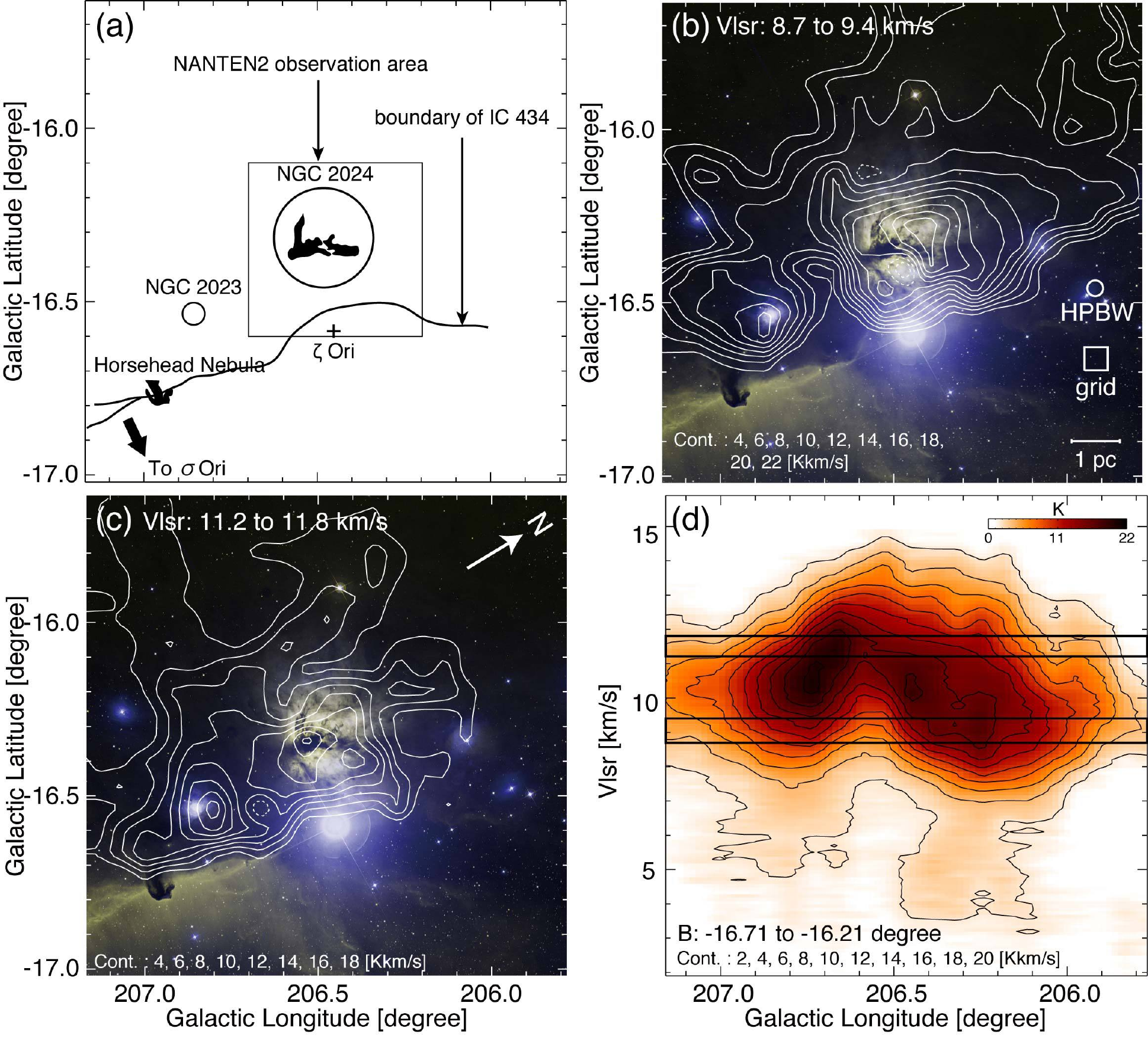}
   \end{center}
\caption{(a) Schematic image of optical features toward the NGC~2023, 2024 region. (b--c) Three color composite image of two optical bands (red, green: DSS2 red, blue: DSS2 blue; STScI Digitized Sky Survey, 1993, 1994, AURA, Inc.) overlaid with contours of \twelvecol ~integrated intensity of the blue-shifted cloud at 8.7 -- 9.4 \kms ~(b) and the red-shifted cloud at 11.2 -- 11.8 \kms ~(c). The contour levels are indicated at the bottom of the Figures. (d) Position-velocity diagram of \twelvecol ~emission integrating -16$\fdg$71 $\le b \le$ -16$\fdg$21. The horizontal lines show the velocity range for the red-shifted cloud and the blue-shifted cloud.}
\label{overview}
\end{figure*}

Figures 1a show schematic image of objects seen in optical wavelengths toward the region including NGC~2023 and 2024. NGC~2023 and 2024 are located near the edge of \htwo ~region, IC434. The black box indicates the area observed by the NANTEN2.
Figures~1b and 1c show integrated intensity distributions of the \twelvecol ~emission toward the same region at 4 \arcmin ~grid obtained with the NANTEN telescope \citep{miz04} superposed on optical images obtained by DSS2.
These two Figures in different integrating velocity range, 8.7 -- 9.4 \kms ~and 11.2 -- 11.8 \kms, present apparently distinct spatial distributions. The higher velocity component exhibits a clear spatial correspondence with the boundary of IC~434, which is a \htwo ~region ionized by $\sigma$-Ori, indicating the association between them.
Figure~1d is a longitude-velocity diagram integrated in latitude. The diagram indicates that there are two peaks shown by a dark color. One is a cloud toward NGC~2023 having \vlsr $\gtrsim$ 10 \kms ~and another is a cloud toward NGC~2024 having \vlsr $\lesssim$ 10 \kms. Gas distributions of these two components correspond to Figure~1c and 1b. There are also diffuse emissions at \vlsr $\sim$5 \kms.

\begin{figure*}
 \begin{center}
  \includegraphics[width=14cm]{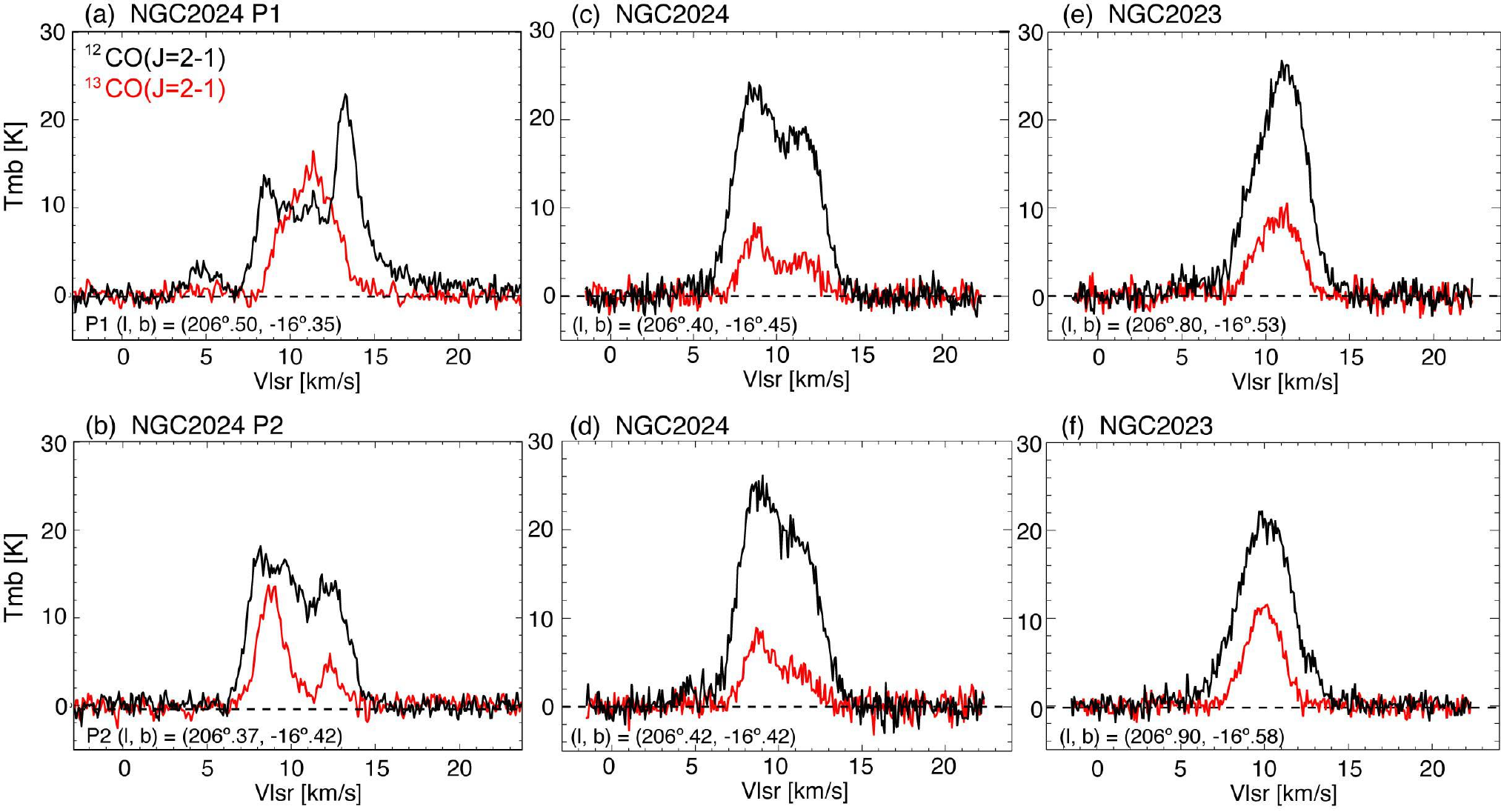} 
 \end{center}
\caption{CO($J$=2--1) spectra at four positions in the NGC~2024 region (a--d) and the NGC~2023 region(e, f). The horizontal and vertical axis indicate a radial velocity and main-beam temperature. \twelvecoh ~and \thirteencoh ~are plotted in black and red lines, respectively.}
\label{profiles}
\end{figure*}

Figures~2a--d show typical \twelveco ~and \thirteencoh ~profiles toward four positions in NGC~2024 and two positions in NGC~2023 obtained from \citet{nis15}. 
Positions P1 and P2, which corresponds Figures~2a and 2b are explained in Section 3.2.
The \twelveco ~emission is heavily saturated with self-absorption in Figure~2a, while \thirteenco ~showing a single peak seems not to have significant saturations.

Indeed some positions in the NGC~2024 region in \twelveco ~apparently show two velocity components due to self-absorption such as seen in Figure~2a, but the region still has two distinct velocity components in some positions as can be seen in Figure~2b--2d. The positions of Figure~2c and 2d do not show any significant emissions in the map obtained by dense gas tracers (e.g., \cite{ike09}), on the other hand \thirteencoh ~data show low-intensity features with two peaks at Vlsr $\sim$ 8.5 and 11.5 km/s.
Also, the catalogue of \citet{ike09} lists three H$^{13}$CO$^{+}$ cores having velocities of 8.8 and 9.2 \kms ~toward P2 (see also \thirteencoh ~spectrum in Figure~2b), but no core having a velocity of $\sim$11.5 \kms.
These indicate that there are two independent velocity features at \vlsr ~$\sim$8.5 and 11.5 \kms. Therefore, the two velocity components shown in Figure~1b--1d are not merely caused by self-absorption but existing two clouds. This also implies that \thirteencoh ~is a very useful tracer in wide density-range of molecular gas toward NGC2024, although this line has some self-absorption toward P1.
If we assume $T_{\rm ex}$ = 30 K, the typical optical depth $\tau (v)$ of \thirteencoh ~toward NGC~2024 is calculated to be $\le$ 0.5 over 95 $\%$ of the whole region, and only a small number of pixels, which have high brightness temperatures $\gtrsim$ 15 K (in the vicinity of P1) have higher optical depths beyond 1 (see Appendix for details).
Therefore, in the present paper, we use only \thirteenco, which is not optically-thick for the most part and traces gas distribution better without self-absorption.

Figures~2c and 2d are line profiles toward two CO intensity peaks in NGC~2023 in Figures~1b and 1c. Both show no significant self-absorption and apparently exhibit a single component. It is possible that two velocity components are merged as the single component in NGC~2023. NGC~2023 is a smaller and younger system compared to NGC~2024 and thus, the angular resolution of the NANTEN telescope is not enough to resolve the two velocity components, although a weak emission is still distinguished at \vlsr $\sim$13 \kms ~in Figure~2d.
Only \twelveco ~spectrum in Figure~2a show a low intensity feature at \vlsr$\sim$5 \kms.
This velocity component was reported to be an independent component associated with Orion B \citep{pet17}. However, the intensity of the velocity component is very law and the distribution is diffuse and thus, this may not play a significant role in the current star formation. Therefore, we do not take into account this component in the following analyses and discussion. 

\subsection{Molecular gas toward NGC~2024}
\begin{figure*}
 \begin{center}
  \includegraphics[width=15cm]{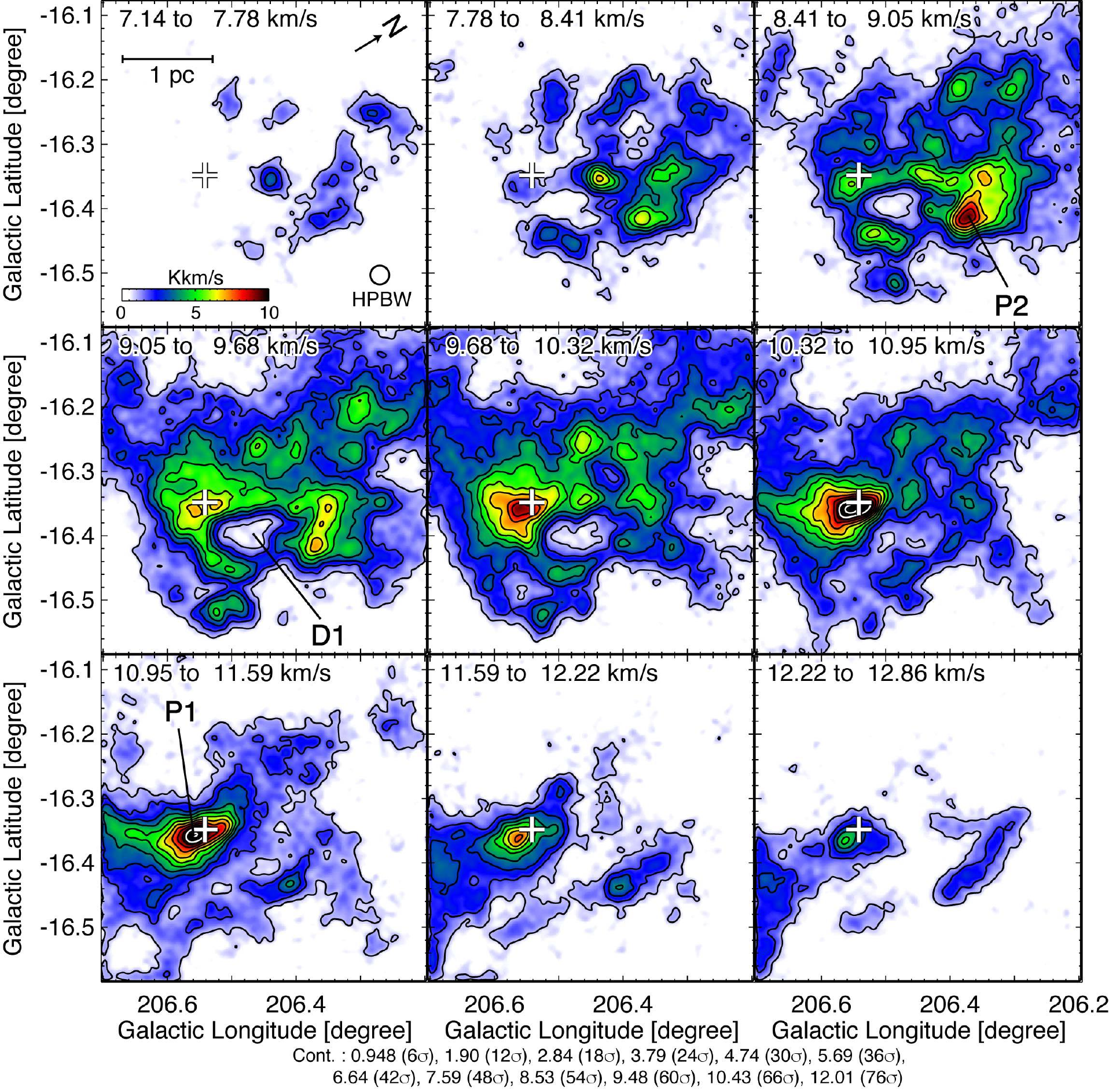} 
 \end{center}
\caption{Velocity-channel distributions of the \thirteencoh ~emission toward NGC~2024 obtained with NANTEN2. The contour levels are indicated at the bottom of the Figure. The cross depicts the position of IRS2b. P1, P2 and D1 show the primary peak intensity ($l$, $b$)=(206\fdg50, -16\fdg35), secondary peak intensity ($l$, $b$)=(206\fdg37, -16\fdg42) and intensity depression ($l$, $b$)=(206\fdg46, -16\fdg40) in \thirteencoh.}
\label{lbch}
\end{figure*}

Figure~3 shows the velocity channel distribution of the \thirteencoh ~emission obtained from new NANTEN2 observations. We found that the primary peak P1 in 9.05 -- 12.86 \kms ~at ($l$, $b$)=(206\fdg50, -16\fdg35) and the secondary peak P2 in 8.41 -- 9.68 \kms ~at ($l$, $b$)=(206\fdg37, -16\fdg42). The two positions show integrated \thirteencoh ~intensity greater than 8 \kkms ~in Figure~3. We also found a marked intensity depression D1 in 8.41 -- 10.32 \kms ~peaked at ($l$, $b$)=(206\fdg47, -16\fdg40). Toward P1, there is the ionizing source of NGC~2024, IRS2b \citep{bik03}. P1 is most likely associated with the source and still might be forming high-mass stars in it \citep{meg12}.

\begin{figure*}
 \begin{center}
  \includegraphics[width=17cm]{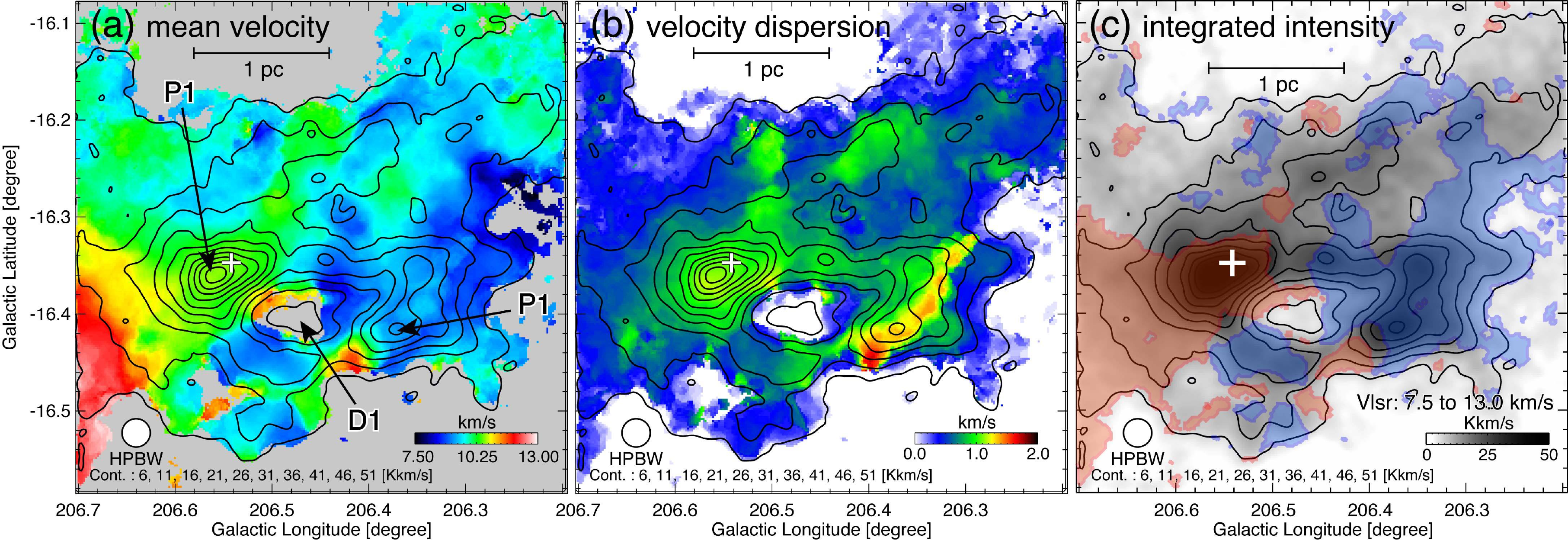} 
 \end{center}
  \caption{(a) Intensity-weighted mean velocity map (1st moment) of the \thirteencoh ~emission in the range of 7.5 $\le \vlsr \le$ 13.00 \kms. (b) Intensity-weighted standard deviation velocity map (The square root of the 2nd moment) of the \thirteencoh ~emission in the range of 7.5 $\le \vlsr \le$ 13.00 \kms. (c) Integrated intensity distribution of \thirteencoh. The red and blue transparent colors correspond pixels with the value of 1st moment beyond 10.3 \kms ~and under 9.5 \kms, respectively. Contours and white cross indicate distribution of integrated intensity over 7.5 $\le \vlsr \le$ 13.00 \kms (0th moment) and position of IRS2b, respectively in all Figures.}\label{}
\label{mommap}
\end{figure*}

These two peaks and the depression with distinct velocity ranges are clearly seen in the 1st moment map of the \thirteencoh ~emission shown in Figure~4a. Figures~4a and ~4b are intensity-weighted mean velocity and standard deviation velocity distributions, respectively with the contours of integrated intensity (0th moment). We used only voxels with enough high significance levels ($\ge$ 6$\sigma$) to calculate the moments in order to reduce effects of noise fluctuations.
Figure~4a shows a systematic velocity gradient: the clouds with the larger galactic longitudes (roughly 206\fdg5 to 206\fdg7) including P1 are red-shifted, and the clouds with the smaller galactic longitudes (roughly 206\fdg2 to 206\fdg5) including P2 are blue-shifted. 

We hereafter refer to these clouds as a red cloud and blue cloud, respectively.
Velocities of these two clouds are generally consistent with those of two clouds shown in the global structure of NGC~2023 / NGC~2024 region (Figure~1b and 1c). 
Therefore, the velocity gradient is not a velocity pattern in a single cloud but the velocity originated from two independent clouds with different velocities.
P1 is included in the red cloud but the velocity of P1 is relatively lower and close to that of the blue cloud (Figure~4a).

Using these two moment maps, we here define the velocity ranges for these two clouds. From Figure~4a, the red cloud shows velocity greater then 10.3 \kms ~corresponding to the light green to the white, and the blue cloud shows velocity less than 9.5 \kms ~corresponding to the blue to the black.
Then, we first defined areas of the red and blues clouds as pixels with $\ge$ 10.3 \kms ~and $\le$ 9.5 \kms ~, respectively. These areas are indicated as the red and blue transparent colors in Figure~4c.

Next, we calculated an averaged-mean velocities and averaged-standard deviation velocities within the defined areas by using two moment maps (Figures~4a and 4b). Here we adopt the averaged-mean velocities and the averaged-standard deviation velocities as the center velocity ($V_{center}$) and the velocity width ($dV$) of a cloud, respectively and define the cloud's representative velocity range as $V_{center} \pm dV$. $V_{center}$, $dV$, velocity range for each cloud are summarized in table~\ref{phys}. The above method of the velocity definition was introduced in \citep{eno19} and they properly identified clouds even in more complicated region, namely the Galactic Center including more than four velocity features in the line-of-sight. The validity for this method in the NGC~2024 region is described in Appendix 2 in detail. 

\begin{table*}[!htbp]
\tbl{Physical parameters of the molecular clouds toward NGC~2024}{%
\begin{tabular}{lccccc} 
\hline\noalign{\vskip3pt}
      cloud name & \vlsr ~[\kms] & \vcen ~[\kms]  & \dv ~[\kms] & \nhtwo ~(peak) [$\times$10$^{22}$ cm$^{-2}$] & Mass [$\times$10$^{2}$ \msun] \\
\hline\noalign{\vskip3pt} 
    blue cloud &  8.5 -- 9.7 & 9.1 & 0.58 & 1.2 & 2.8  \\
    red cloud & 10.4 -- 11.6 & 11.0 & 0.61 & 2.4 & 2.3 \\
\hline\noalign{\vskip3pt} 
\end{tabular}}
\label{phys}
\begin{tabnote}
\hangindent6pt\noindent
Note. --- Col.1: Names of clouds. Col.2: Velocity ranges. Col.3: Peak velocities derived from the 1st moment map. Col.4: Velocity line widths derived from the 2nd moment map. Col.5: Maximum molecular column density toward each cloud. Col.6: Molecular mass.\\
\end{tabnote}
\end{table*}

Figure~4b shows a correlation between velocity dispersions (color) and integrated intensities (contours) other than a filamentary region in the vicinity of P2. The filament indicates higher value of $\ge$ 1.2 \kms ~due to a overlap of two clouds in the line-of-sight (see Figure~3).

\begin{figure*}
 \begin{center}
  \includegraphics[width=14cm]{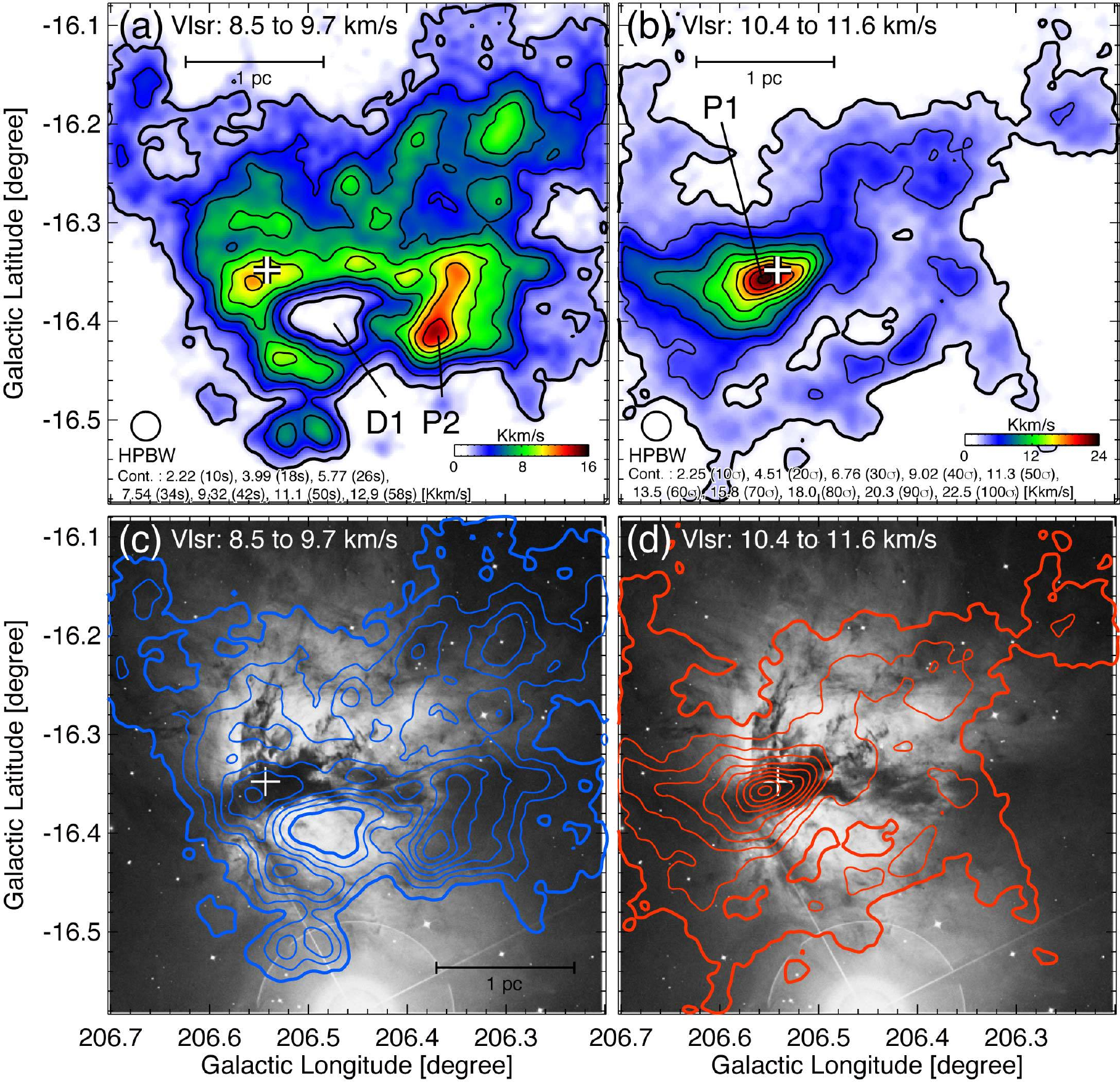} 
 \end{center}
\caption{(a, b) Integrated intensity distributions of \thirteencoh ~with the velocity ranges defined as the blue cloud (8.5 $\le \vlsr \le$ 9.7 \kms) and the red cloud (10.4 $\le \vlsr \le$ 11.6 \kms). Contour levels are indicated at the bottom of each Figure. (c, d) The same contours as Figure 5a and 5b but superposed on optical images of DSS2 red. The white cross indicates the position of IRS2b.}
\label{rb}
\end{figure*}

Figures~5a and 5b are integrated intensity distributions of the blue and red clouds, respectively using integration velocity ranges defined above. The blue cloud is a diffuse cloud extending $\sim$2 pc which has a significant hole (D1) at the bottom center of it. On the other hand, the red cloud mainly consists of a small $\sim$1-pc clump (P1). Figures~5c and 5d are superpositions of contours of the red and blue clouds on DSS2 r band images.

The DSS2 image shows a major optical dark lane across the \htwo ~region. Some of the additional minor dark lanes have a tilt of $\sim70^{\circ}$ to the major lane. The blue cloud shows better correlation with the dark lanes, suggesting that the cloud is located on the near side of the H{\sc ii} region (Figure~5c). The red cloud is most likely associated with the exciting star (IRS2b) and thus lies inside or behind the \htwo ~region as shown by its poor correlation with the dark lanes (Figure~5b).

The association between the blue cloud and the \htwo ~region is not clear only above observational evidence. However since the column density of the cloud is not so low and it is unnatural to consider that such a relatively dense gas are located between the known nearest clouds, the Orion clouds, and us in the same line-of-sight by chance, it is plausible to consider that the blue cloud is a part of the NGC~2024 system and associated with the \htwo ~region.

\subsection{physical parameters of molecular clouds}
The column densities and masses of the red and blue clouds are derived from our \thirteencoh ~data as follows: We first assumed that the excitation temperature \tex ~of all the \thirteenco ~clouds to be 30 K (LTE approximation) because the maximum value of \tmb ~is $\sim$20 K in \thirteencoh ~and a estimated maximum value of \tex ~from \twelvecol ~is $\sim$30 K. The equivalent brightness temperature $J(T)$ is described as
\begin{equation}
  J(T) = \frac{h\nu}{\kb}~[~\exp(\frac{h\nu}{\kb T}) - 1~]^{-1}.
\end{equation}
where $h$, k$_{B}$, and $\nu$ ~are Planck constant, Boltzman constant, and the frequency, respectively. From the radiative transfer equation, we obtain optical depth $\tau_{\nu}$ as follows,
\begin{equation}
  \tau (v) = -\ln(~1~-~\frac{\tmb}{J(\tex)~-~J(\tbg)}~).
\end{equation}
Using the $\tau_{\nu}$, column density $N$ of the molecule can be calculated as
\begin{equation}
  N = \sum_{V} ~\tau (v)~\Delta v~\frac{3\kb\tex}{4\pi^{3}\nu\mu^{2}}~\exp(\frac{h\nu J}{2\kb\tex}) ~\frac{1}{1 - \exp({-\frac{h\nu}{\kb\tex}})}.
\end{equation}
where $V$, $\Delta v$, $\mu$, and $J$ are the velocity range of a cloud, the velocity resolution of the data, the electric dipole moment of the molecule, and the rotational transition level, respectively. Substituting $\kb$ = 1.38 $\times~10^{-16}$ (erg/K), \tex ~= 30 (K), $\nu$ = 2.20 $\times~10^{11}$ (Hz), $\mu$ = 1.10 $\times~10^{-19}$ (esu~cm), $h$ = 6.63 $\times~10^{-27}$ (erg~s), $J$ = 1 into equation (3) gives below
\begin{equation}
  \nco = 1.508 \times 10^{16} ~\sum_{V} ~\tau (v)~\Delta v.
\end{equation}
 Assuming ratio between \nhtwo ~and \nco ~of 5 $\times~10^5$ \citep{dic78}, \nhtwo ~is estimated to be
\begin{equation}
  \nhtwo = 7.54 \times 10^{21} ~\sum_{V} ~\tau (v)~\Delta v.
\end{equation}
 
The velocity ranges of two clouds which we defined before well represent major features of each clouds (Figures~5a and 5b). However, diffuse emissions are still extending over these velocity ranges (see Figures~2a and 2b). Therefore, we here adopt 7.50 $\le ~\vlsr ~\le$ 13.00 \kms ~as the velocity range for deriving column densities and masses of two clouds ($V$ in equation (3)--(5)). We also use red and blue distributions in Figure~4c as the definition of the red and blue clouds in space. Applying above definitions in space and velocity, we obtained typical/peak column densities of the red and blue clouds to be 7.7/240 $\times ~10^{20}$, 9.3/190 $\times ~10^{20}$ cm$^{-2}$, respectively. Molecular mass is estimated following equation:
\begin{equation}
  M = \mum \mpro \sum_{i} ~[d^2 \Omega \nhtwo_{,i}].
\end{equation}
where $\mum$, $\mpro$, $d$, $\Omega$ and $\nhtwo_{,i}$ are the mean molecular weight, proton mass, distance, solid angle subtended a pixel, and column density of molecular hydrogen for the i-th pixel, respectively. We assume a helium abundance of 20 $\%$, which corresponds to $\mu_{w}$ = 2.8, and we take $d$ = 410 pc and then get masses of the red and blues clouds of $\sim$230 and $\sim$280 \msun, respectively. These values are not so different compared to that of previous study ($\sim$ a few hundred \msun; \cite{kra96}).
These physical parameters are summarized in table~\ref{phys}.

The column density of P1 derived our \thirteencoh ~data are not high ($\sim$2.4 $\times 10^{22}$ cm$^{-2}$) compared to the column density derived from $Herschel$ observations ($\sim$1--2 $\times$ 10$^{23}$ cm$^{-2}$; see Figure~9 of \cite{gra17}). Therefore, our values might be lower limits caused by a beam dilution effect due to the coarser angular resolution. Observations with higher angular resolution are required to clarify this.

\section{Discussion}
In the previous section, we presented not only a global distribution and properties of molecular gas but also gas distribution with finer angular resolution obtained by our new comprehensive \thirteencoh ~observations. The results show that there are prominent molecular features; two intensity peaks P1, P2 and a depression D1.
P1 is a major feature of the one of associated clouds, the red cloud, with NGC~2024. P2 and D1 are parts of another associated cloud, the blue cloud, located in front of the \htwo ~region. 
 
 In the present section, we focus on the origin of marked features and discuss a possible high-mass star formation scenario in this region.
  
\subsection{Complimentary distribution}
D1 is an unusual marked hole with sharp intensity gradient toward its surroundings. The interior of it shows almost no significant \thirteenco ~emissions. It is hard to explain that the structure was created by chance in a process of the self-gravitational evolution of the blue cloud. Such a hole is usually observed toward surroundings of a high-mass star(s) as a result of ionization by its UV radiation. However, there is no known high-mass stars interior of D1 (e.g. \cite{meg12}).

Another mechanism to create such a hole is cloud-cloud collision (CCC). If a CCC is operating, we expect a complementary distribution between the hole and a clump, which hollowed out the blue cloud and created D1. Therefore, assuming D1 was created by a CCC, another independent cloud with a different velocity from the blue cloud is required. The only candidate is the red cloud. By eye inspection, we recognize a complementarity between P1 and D1 (see Figures~5a and 5b).

\begin{figure*}
 \begin{center}
  \includegraphics[width=8cm]{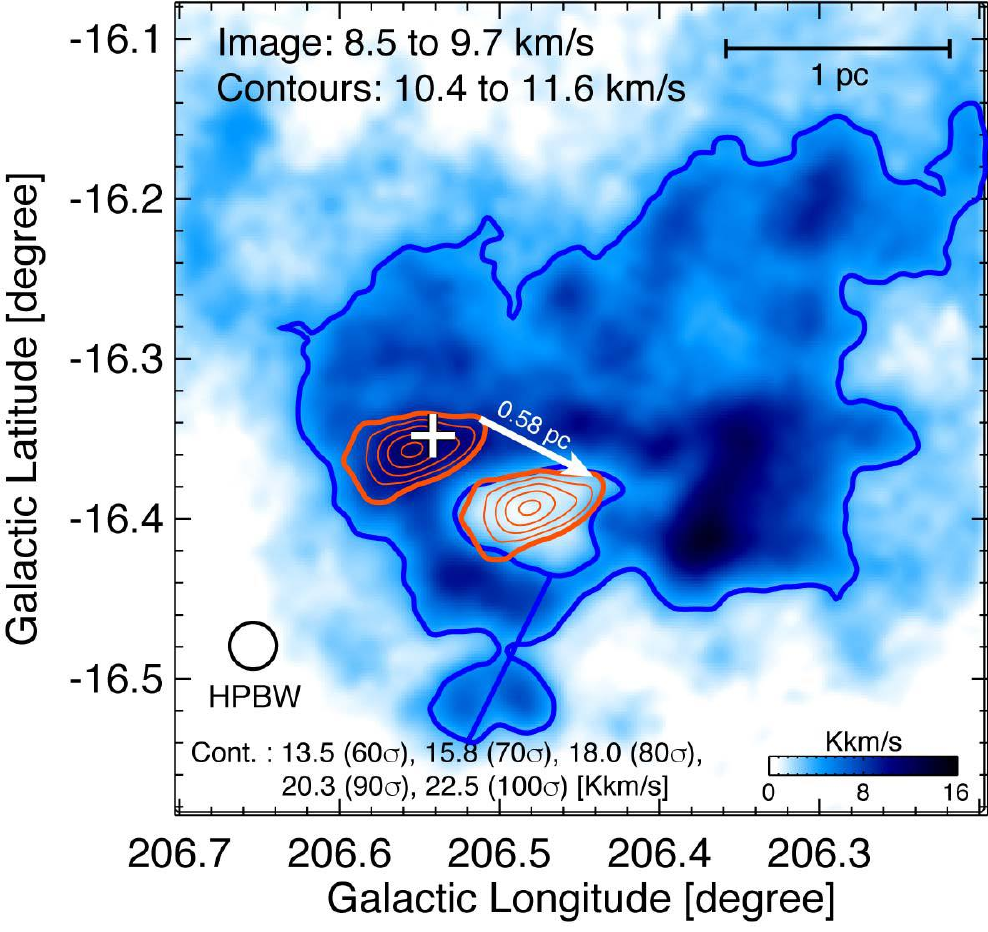} 
 \end{center}
  \caption{Integrated intensity distribution of the red cloud in \thirteencoh. The thin and thick orange contours indicate distribution of P1 in the original position and displaced position $\sim$0.6 pc to south-west. The white arrow indicates the displacement vector.}\label{}
\label{displace}
\end{figure*}

Figure~6 shows that distribution of the blue cloud in blue and P1 in orange contours.
We here increase the lowest level of contours for P1 to 60 $\sigma$ (= 13.5 \kkms) because P1 is overlapped with a filamentary cloud toward ($l$, $b$, $v$) = (206\fdg6 to 206\fdg7, -16\fdg4 to -16\fdg3, 10.32 to 11.59 \kms) in the line-of-sight and its distribution is contaminated by the filamentary cloud below $\sim$13 \kkms.
We found that P1 well fits to D1 if they are displaced $\sim$0.6 pc to the south-west with the angle of  26.57 degree clockwise from the direction of the galactic longitude indicated by a white arrow.
Such a displacement is an observational signature of colliding clouds \citep{fuk18a,fuk18b}.
According to \citep{fuk18b}, the complementarity between the clump in the smaller cloud and the hole in the larger cloud is easily weaken by projection effect and the shapes of these two do not necessarily coincide perfectly.

\subsection{Possibility of a cloud-cloud collision in NGC~2024}
We mentioned that a CCC between the red and blue clouds is a plausible mechanism to create D1. We examine a further applicability of the CCC scenario below.

\citet{fuk18b} summarized observational signatures of high-mass star formation under triggering by cloud-cloud collision as follows;
\begin{enumerate}
\renewcommand{\labelenumi}{\roman{enumi})}
\item Two clouds with supersonic velocity separation associated with young high-mass star(s),  
\item complementary spatial distribution between the two clouds, and 
\item bridge feature connecting the two clouds in velocity.
\end{enumerate}

According to the theoretical simulations of \citet{tak14} and the synthetic observations by \citet{fuk18b}, an intensity depression is produced by a cloud-cloud collision where a smaller cloud creates a hole in a larger cloud.
The displacement between the small cloud and the hole reflects a tilt angle of the relative cloud motion to the line of sight, and synthetic observations of two colliding clouds based on numerical simulations provide details of the collisional interaction between the small cloud and the hole \citep{fuk18b}, as described in Section 4.2.
The numerical simulation show the collisional front is rapidly compressed and massive clumps which will grow up progenitor of high-mass stars are formed in it. Once two clouds begin to collide, they produce a bridge feature or a V-shaped structure in the position-velocity digram due to the exchange of momenta. Recently, thanks to galactic plane surveys with enough high angular resolution in CO such as FUGIN \citep{ume17} and Mopra \citep{bra18}, more than 50 galactic \htwo ~regions and clusters that have above signatures triggered by CCC are reported (e.g., \cite{eno18, hay18, san18, dew19}).

Now, we have two clouds (the red and blue) with supersonic velocity separation of $\sim$2 \kms ~associating a high-mass star IRS2b and found these two clouds exhibit complementary distribution with a certain displacement.

\begin{figure*}
 \begin{center}
  \includegraphics[width=16cm]{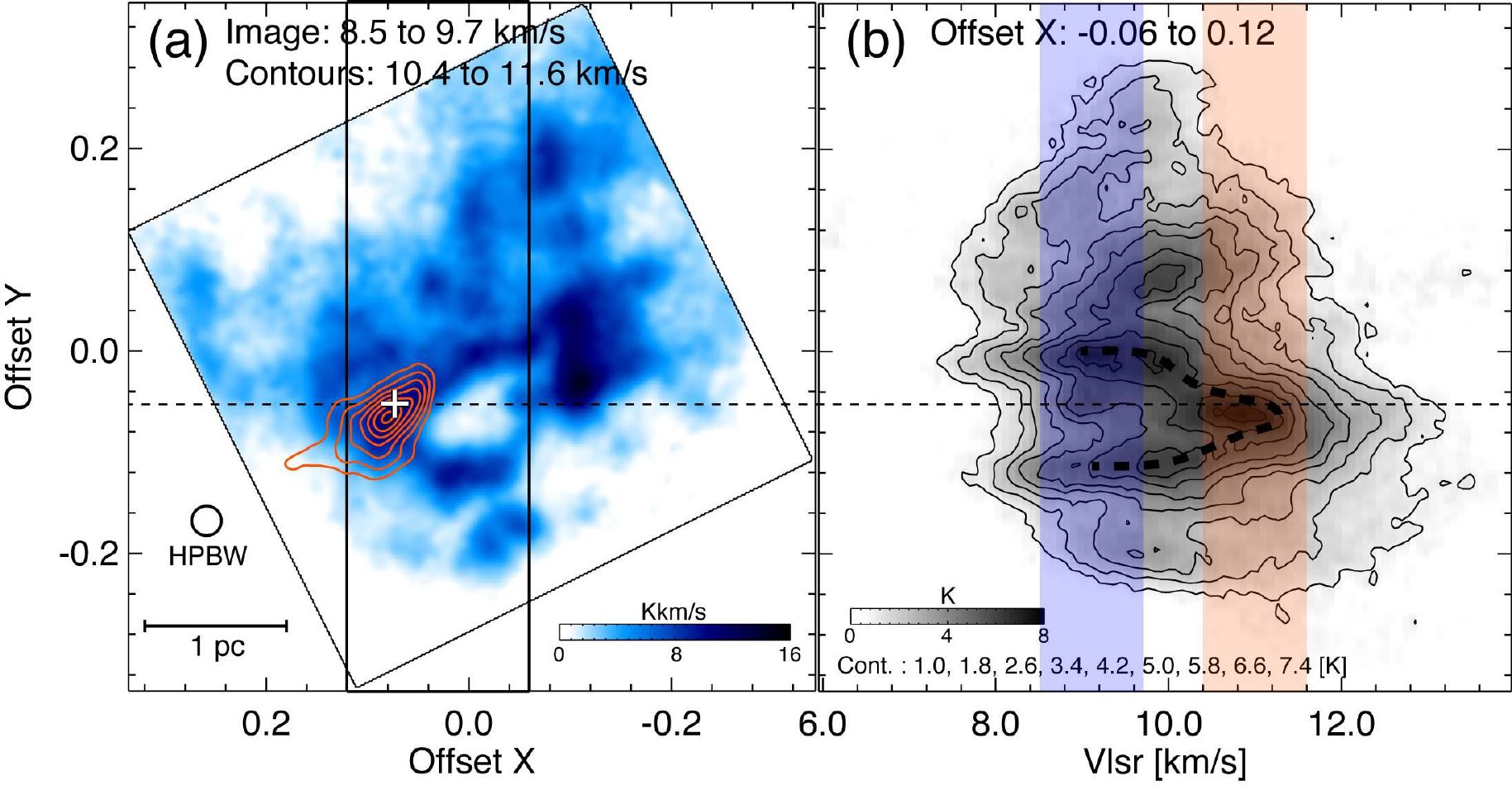} 
 \end{center}
\caption{(a) The \thirteencoh ~distribution of red and blue clouds in the offset X-Y coordinate. The offset XY coordinate is defined by rotating the galactic coordinate counterclockwise by 26.57 degree. (b) Offset X--velocity diagram of the NGC~2024 region in \thirteencoh. Integration range in the offset Y is indicated in the top of the Figure and black lines in Figure 7a. Velocity ranges of the red and blue clouds are indicated as red and blue transparent belts, respectively. The white cross and thin black dashed line indicate the position and the offset Y coordinate of the exciting star, IRS2b.}
\label{vshape}
\end{figure*}

Figures~7a and 7b are a integrated intensity distribution and the position-velocity diagram of the red and blue clouds in the offset X-Y coordinate. The coordinate was defined by rotating the galactic coordinate counterclockwise to the angle given by the displacement vector (= 26.57 degree) indicated by the white arrow in Figure~6. The position-velocity diagram shows a rotated V-shape connecting the red and blue clouds indicated by a thick black dashed line. The toe of the V shape corresponds to the location of IRS2b.

In conclusion, we suggest that a CCC happened in NGC~2024.
It could be argued alternatively that the two velocities are due to acceleration by the late O/early B-type stars and not a cause of a CCC. The cloud velocity and dispersion, however, show no systematic enhancement or variation toward or correlated with the O/B-type stars (see Figure~7b), which is not consistent with a dominant dynamical effect by the stars. The velocity field which does not show particular variation toward the stars is odd, if the stars were the major source of the cloud momentum. So, we do not consider the stellar acceleration, and explore the CCC as a possible scenario in the present paper.

\subsection{Triggered star formation by a cloud-cloud collision model}
\begin{figure*}
 \begin{center}
  \includegraphics[width=8cm]{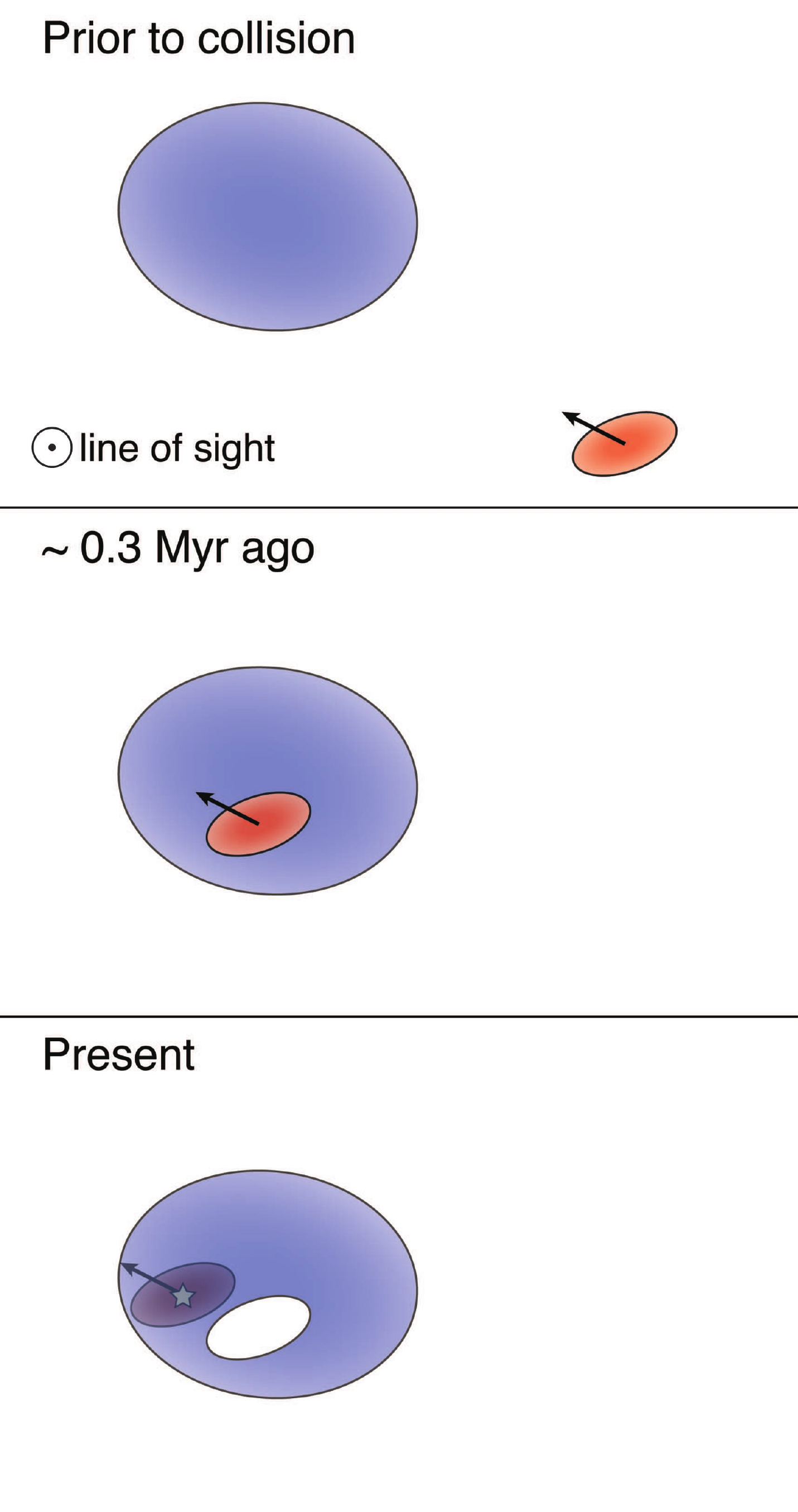} 
 \end{center}
\caption{Schematic image of the CCC in the NGC~2024 region in the sky view.}
\label{schematic}
\end{figure*}
Based on the above results and discussion we hypothesize a collision between the two clouds triggered formation of the late O/early B stars in NGC~2024. For simplicity, if we assume a tilt angle $\theta$ of $45^{\circ}$ to the line-of-sight, the cloud relative velocity and the displacement are 1.9 $\times ~\sqrt{2}$ $\simeq$ 2.7 \kms ~and 0.6 $\times ~\sqrt{2}$ $\simeq$ 0.8 pc, respectively. The timescale of the collision is then estimated to be 3 $\times$ 10$^5$ yrs from a ratio 0.8 pc/2.7 \kms. A schematic image of the collision in the sky view is shown in Figure~8.

The time scale is consistent with a very small age less than Myr of the young stars in NGC~2024 \citep{ali95,mey96}. If we assume a tilt larger than $60^{\circ}$ the velocity becomes 4 \kms ~or more, whereas an assumption on the tilt angle does not significantly alter the timescale. The location of the blue cloud in the foreground of the \htwo ~region is consistent with an epoch after the collision.
According to \citet{fuk20}, the compressed layer in a CCC develops dense clumps having the mass accretion rate of 10$^{-4}$ \msun/yr within 0.3 Myr, if the initial gas density is 300 cm$^{-3}$. Assuming that the dense clumps maintain the mass accretion rate of 10$^{-4}$ \msun/yr, 0.15 Myr is required to form a high-mass star of 15 \msun. It is almost comparable to the time scale of the collision in NGC~2024 of 0.3 Myr estimated by the displacement within the error range of 50 \%.
If we assume that the density of P1 before the collision (initial density of P1) is as same as that of current P2, we can estimate the initial density to be the column density $\div$ size = 1.2 $\times$ 10$^{22}$ cm$^{-2}$ $\div$ 0.25 pc $\sim$ 1600 cm$^{-3}$. \citet{fuk20} suggested that the required time to generate dense clumps is shorter if the initial density is higher than 300 cm$^{-3}$. Therefore, it is possible that the error is mainly responsible for the higher initial density in the NGC 2024 region.

\begin{figure*}
 \begin{center}
  \includegraphics[width=12cm]{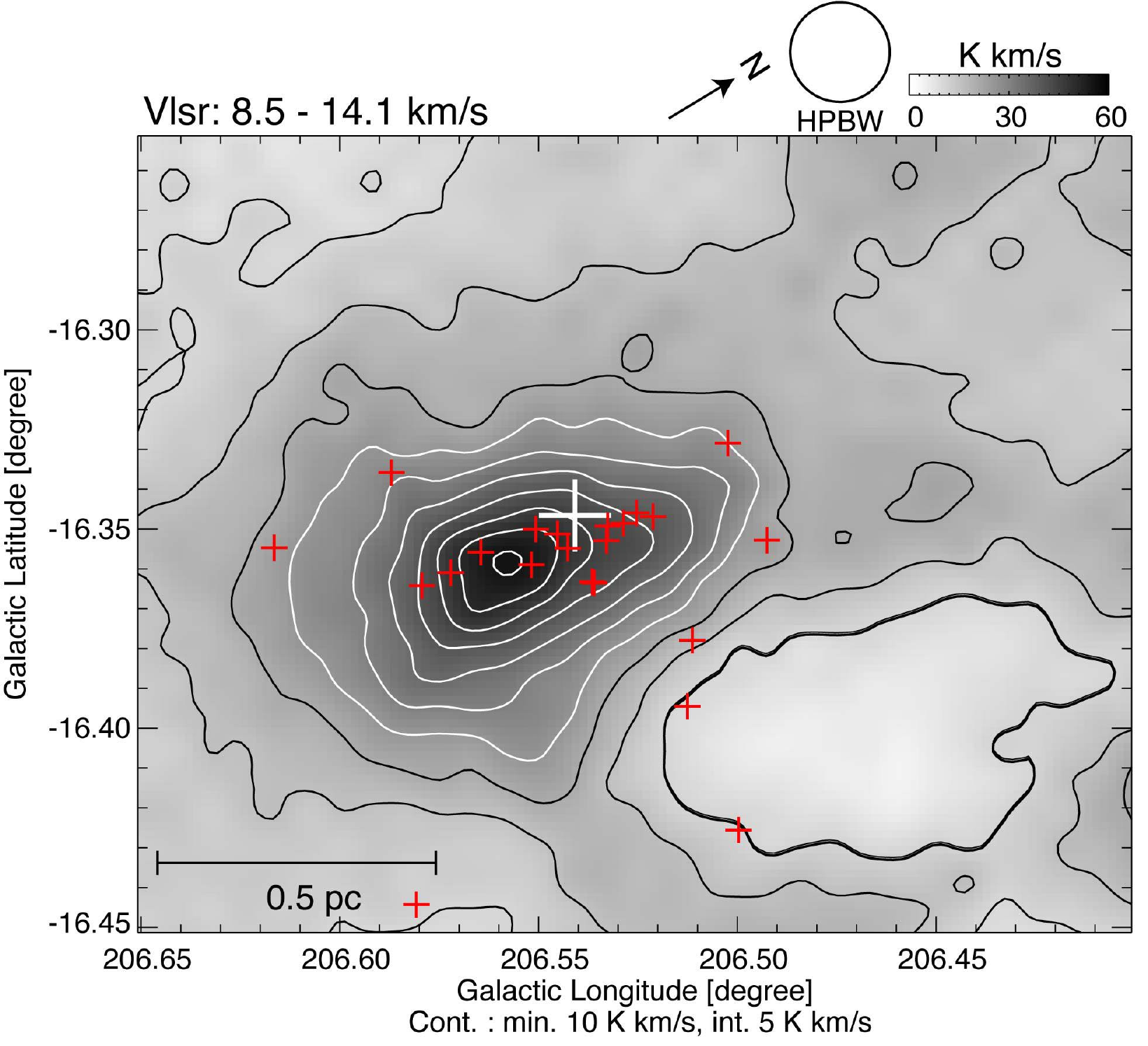} 
 \end{center}
\caption{The \thirteencoh ~distribution of the P1 with the velocity integration range of 8.5 $\le \vlsr \le$ 14.1 \kms. The white and red crosses indicate the positions of IRS2b and protostars \citep{meg12}.}
\label{sf}
\end{figure*}

Figure~9 shows the \thirteenco ~clump toward the primary peak P1; this indicates a strong concentration of $\sim$ 20 protostars \citep{meg12}, which corresponds to an O-type star IRS2b and probably young B stars \citep{get14,bik03,ski03}, toward the \thirteenco ~peak P1, showing that the \thirteenco ~clump is forming high-mass stars actively.

In the present scenario, the relative velocity of the clouds is pre-determined in the Galactic environment and the collision is by chance. The gravity of the system is actually not dominant for a set of relevant parameters; the velocity in gravitational balance with an observed cloud mass of 200 \msun ~and a radius of 1 pc is 1.5 \kms, marginally less than the projection-corrected velocity above.

In the scenario, the major collision took place toward P1 in the red cloud, which was impacted by the blue cloud on the far side of the red cloud.
P1 has a high column density of more than 10$^{23}$ cm$^{-2}$ and $\sim$ 20 protostars are forming at present \citep{meg12}. 
The interface layer between the two colliding clouds becomes highly turbulent due to the clumpy distribution in the cloud prior to the collision. The turbulence amplifies the magnetic field according to the numerical simulations of cloud-cloud collision \citep{ino13}. The combined contribution of the turbulence and magnetic field realizes high-mass accretion rate of 10$^{-3}$ to 10$^{-4}$ $M \odot$/yr, which allows a protostar to overcome the stellar radiation pressure and to grow in mass. The most high-mass star in NGC~2024 IRS2b having 23 $M \odot$ \citep{bik03} can be formed in a timescale of 1 $\times$ 10$^5$ yrs at an assumed mass accretion rate 2 $\times$ 10$^{-4}$ $M \odot$/yrs. The column density at P2 where the collision have not been took place is about half of that at P1, then if we assume D1 had the column density comparable to that of P2 prior to the collision, the collision increased the column density by a factor of two.

 The heavy obscuration toward the young stars and the sign of disks of the stars \citep{mey96} is consistent with the scenario and do not exclude that the stars are still growing in mass via accretion to become O-type stars eventually in $\sim$10$^5$ yrs, when ionization by O-type stars may halt further mass accretion.

\subsection{NGC~2024 in samples of cloud-cloud collision}
We summarized signatures of a CCC suggested in \citep{fuk18b} in subsection 4.2.
These signatures are however not always detectable in observational data. The numerical simulations (\cite{tak14}, and the synthetic observations by \cite{fuk18a}) indicate that the two clouds often show a single spectral peak instead of two peaks, because the collision mixes the two clouds in velocity as a result of momentum exchange between the two clouds (see Figures 3 and 4 in \cite{fuk18a}).

The trend of a single peak in colliding clouds becomes significant if the projected velocity separation is smaller than the linewidth of the individual clouds. The peak velocity of the merging clouds is governed by the cloud with higher molecular column density \citep{fuk18b}. The collisional dissipation further destroys the two clouds \citep{tor15}, and in addition, the ionization by the formed O-type star(s) disperses the parent molecular gas within $\sim$10 pc of the O-type stars.
As a result, the collision signatures quickly disappear (Figure 4 in \cite{fuk16}). These effects make it difficult to identify the collision signatures and the initial two clouds, and an apparently single cloud does not contradict with CCC.

In NGC~2024, the projected velocity separation between the two clouds is small, only 2 \kms, and we see the complementary distribution (Section 4.1). The two clouds appear to be merged toward P1 which shows large velocity dispersion from 8 to 13 \kms ~at 3 K pc level (Figure~7b). Due to the small duration time, the collisional cloud dissipation is still not significant and we see a massive dense clump toward the $\sim$20 protostars as P1 whose molecular mass is estimated to be 60 $M \odot$ from \thirteencoh ~data by assuming LTE within 0.2 pc.
The absence of early O-type stars makes ionization less effective than in O-type star forming regions. 

\citet{eno19} suggested that there is a column density threshold to trigger formation of high-mass stars by CCC.
The threshold molecular column density in formation of multiple O-type stars is $\sim$10$^{23}$ cm$^{-2}$ and that in a single O-type star 10$^{22}$ cm$^{-2}$. For column density less than these values no O-type stars are formed in CCC, and only collision toward high column-density gas leads to formation of O-/early B-type stars. High-mass star clusters in RCW38, Westerlund 2, NGC~3603, and M42 have molecular column density of $\sim$10$^{23}$ and some authors suggested these clusters are triggered by CCC \citep{fuk16,fur09,fuk14,fuk18b}.
 The high column density in NGC~2024, $\sim$ 10$^{23}$ cm$^{-2}$, is consistent with this threshold, and we infer that even more high-mass stars may form in near future if high-mass star formation is continuing in NGC~2024. 

Both the red and blue clouds discovered by our work is extending to the NGC~2023 region (Figure~1b, 1c). Therefore it is suggested that a CCC is happening also in this region, although spectra at current spatial resolution show only a single component (Figure~2c and 2d).
A further discussion of CCC and high-mass star formation in Orion B will be developed in a forth coming papers on NGC~2023 (Fukui et al. 2020b in preparation) and NGC~2068 / NGC~2071 \citep{fuj19}.

\citet{fuk20} statistically analyzed a MHD simulation of colliding flows taken by \citet{ino13} and concluded that CCC is a mechanism which selectively creates a top-heavy core mass function (CMF), and then serves more high-mass stars compared to other star-formation mechanisms, through comparing the simulation and observations (RCW120, M20, RCW38, W43). They estimated the SFE of only one object (RCW120) to be $\sim$1 \% which is not as high as the typical value in the Galaxy.
This infers that CCC enhances top-heavy CMF with maintaining the typical SFE. We will examine this possibility by analyzing comprehensive CO and infrared data in the Orion region in near future.


\section{Conclusions}
We carried out new observations of \thirteencoh ~transitions in the NGC~2024 region with NANTEN2. These observations cover the whole NGC~2024 region at 0.2 pc resolution. The main results are summarized below. 

\begin{enumerate}
\item Contrary to the previous observations which suggested a single cloud component, we found a possibility that the cloud comprises two velocity components (a red cloud and blue cloud) with complementary distribution. The projected velocity separation of these clouds is $\sim$ 2 \kms. The blue cloud shows good correspondence with several minor dark lanes, indicating that it is on the foreground of the H{\sc ii} region. On the other hand, the red cloud most likely associates with the exciting star IRS2b and is located inside or backside of the \htwo ~region.

\item We found that a displacement of $\sim$0.6 pc in a position angle of $\sim 27^{\circ}$ of the blue cloud produces a good spatial correspondence between an intensity peak in the red cloud and a depression in the blue cloud forming complementary distribution. These two velocity components seem to be merged into a single velocity component.

\item We hypothesize that collision between the two clouds triggered formation of the $\sim$ 20 protostars including IRS2b. The collision timescale is estimated to be 3 $\times$ 10$^{5}$ yrs. The molecular column density toward the \thirteencoh ~peak estimated to be higher than $\sim2 \times 10^{22}$ cm$^{-2}$, possibly $\sim10^{23}$ cm$^{-2}$ , and that in the surrounding regions is $\sim1 \times 10^{22}$ cm$^{-2}$. The collision realized a high-mass accretion rate of 10$^{-4}$ $M \odot$/yrs as shown by the mechanism presented by \citet{ino13}, and triggered the O/early B star formation. The number of high-mass star candidates is consistent with that of the expected number by its high column density.
\end{enumerate}

It is important in future to investigate nearby molecular clouds within a few kpc in order to better establish the role of CCC in O-type star formation.
The coarse resolution CO data in Figures~1b and 1c does not exclude two velocity components, which might suggest a CCC toward NGC2023. It is therefore important to explore a possibility of CCC at higher resolution toward NGC~2023 also. NGC~2024 is a unique object which is extremely young with an age $\lesssim$ $10^5$ yrs among those discovered until now. Thanks to the young age the parent cloud remains unionized and we are able to observe its details including the column density. Higher resolution studies with ALMA will shed a new light on the formation of dense clumps and their mass function in the shock-compressed layer by CCC.

\begin{ack}
We thank (an) anonymous referee(s) for very helpful comments that improved the manuscript.
This work was financially supported Grants-in-Aid for Scientific Research (KAKENHI) of the Japanese society for the Promotion for Science (JSPS; grant number society for 15K17607 and 15H05694). NANTEN2 is an international collaboration of ten universities, Nagoya University, Osaka Prefecture University, University of Cologne, University of Bonn, Seoul National University, University of Chile, University of New South Wales, Macquarie University, University of Sydney, and Zurich Technical University.
\end{ack}

\newpage

\appendix 

\section{Opacity of the \thirteencoh ~line emission}
In this section, we evaluate the validity of the \thirteencoh ~line emission as a tracer of molecular gas associated with NGC~2024. Within the velocity range of the associated clouds (7.5 to 13.0 \kms), we calculated optical depths ($\tau (v)$) of \thirteencoh ~for all voxels by using LTE approximation with \tex ~= 30 K. Voxels without significant line emissions (below 5 sigma = 4.17 K) were flagged in advance.
Figures~10a, and 10b show histograms of the \thirteencoh ~intensity, and $\tau (v)$, respectively. These figures indicate that weak, optically thin emissions dominate in this region. Voxels with $\tau (v)$ $\ge$ 0.5 occupy only 5.3 \% of all voxels. Figures~10c, and 10d show the same plots as figures~10a, and 10b but the vertical axis is the number of voxels times tau, which is proportional to mass. Voxels with $\tau (v)$ $\ge$ 0.5 occupy only 10.3 \% of total mass, means a saturation is not significant (less than 10.3 \%) in this reagion.
\begin{figure*}
 \begin{center}
  \includegraphics[width=14cm]{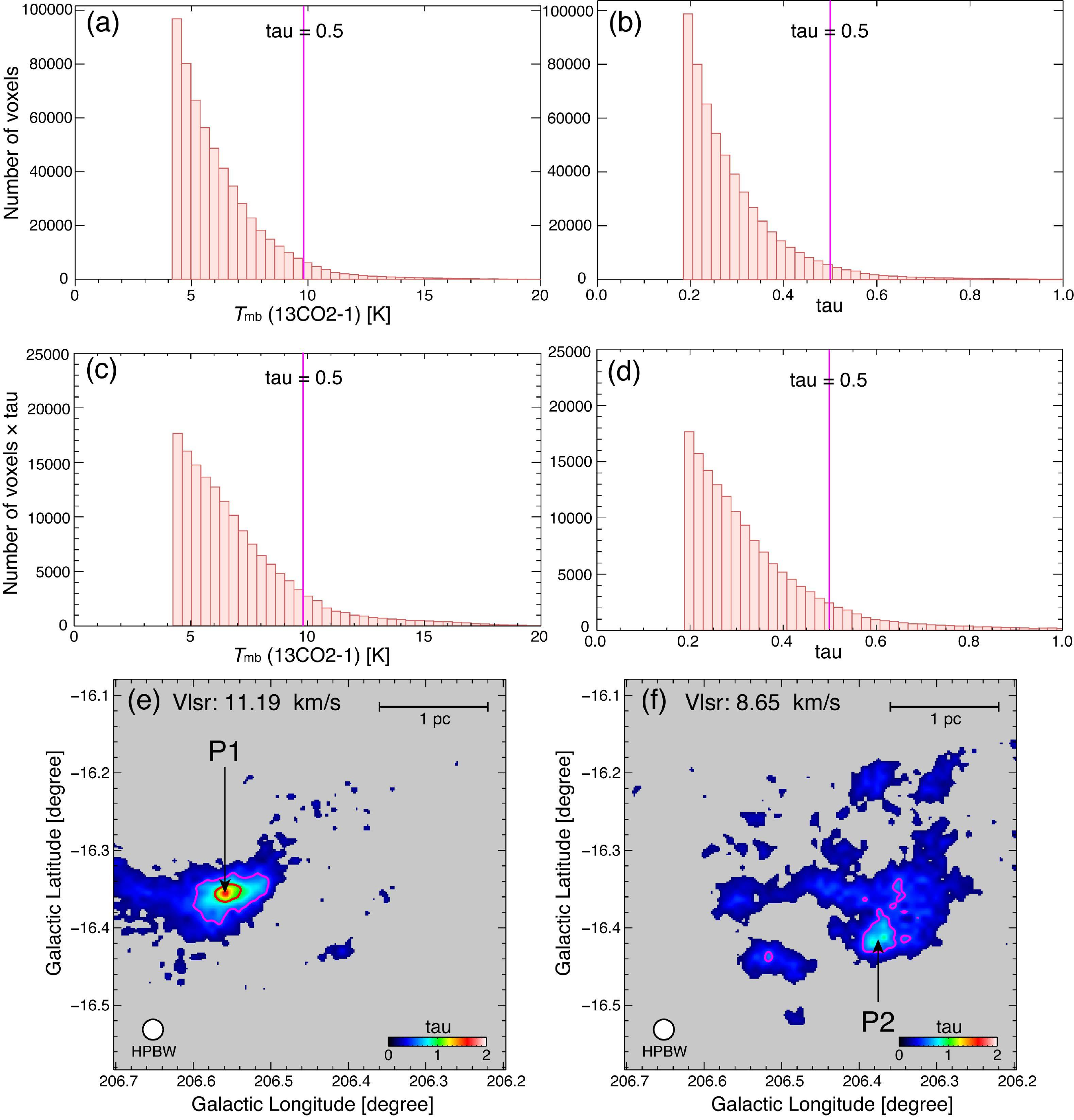} 
 \end{center}
\caption{Histograms of the brightness temperature of \thirteencoh ~line emissions (a), and of optical depths derived from the LTE approximation (b) for all voxels. The magenta line indicates $\tau (v)$ = 0.5. (c, d) The same plots as figures~10a, and 10b but the vertical axis is the number of voxels times tau. (e,f) Distributions of optical depths at the \vlsr ~= 11.19 and 8.65 \kms, respectively. The magenta and red contours correspond $\tau (v)$ = 0.5, and 1.0, respectively.}
\label{tau}
\end{figure*}

Figures~10e, and 10f show spatial distributions of optical depths at two prominent velocities including P1 and P2. P1 has the highest value of $\tau (v)$ beyond 1.0 in the NGC~2024 region. However, the value is not extremely high of $\sim$1.5 and the other molecular gas traced by \thirteencoh ~are obviously optically thin ($\tau (v)$ $<$ 0.5). Such diffuse gas has not been detected by dense gas tracers (e.g., \cite{ike09}).
From this evaluation, we concluded that \thirteencoh ~is a useful tracer for both diffuse/dense gas in the NGC~2024 region even though it has significant absorption at P1.

\section{Velocity-range definition for the two clouds}
We determined velocity ranges for the two associated clouds by using two moment maps (Figures~4a, 4b) in Section 3.2. This method (hereafter moments method) was proposed by \citet{fuk18b} and has been commonly used to identify two clouds (e.g., \cite{eno19}). We here evaluate whether it is useful also in the NGC~2024 region.
As suggested by \citet{fuk18b}, the Orion region has comprehensive two or more velocity components and owing to the proximity of their velocities multiple components are often merged with each other. In the NGC~2024 region, peak velocities of the red and blue clouds vary point by point from $\sim$7 to $\sim$11 \kms ~and from $\sim$9 to $\sim$15 \kms, respectively. This means that any two velocity ranges cannot completely draw distributions of the two clouds.

In order to present the distributions of these two velocity components, we performed a double-gaussian fitting for each line profile in the \thirteencoh ~data cube.
\begin{figure*}
 \begin{center}
  \includegraphics[width=14cm]{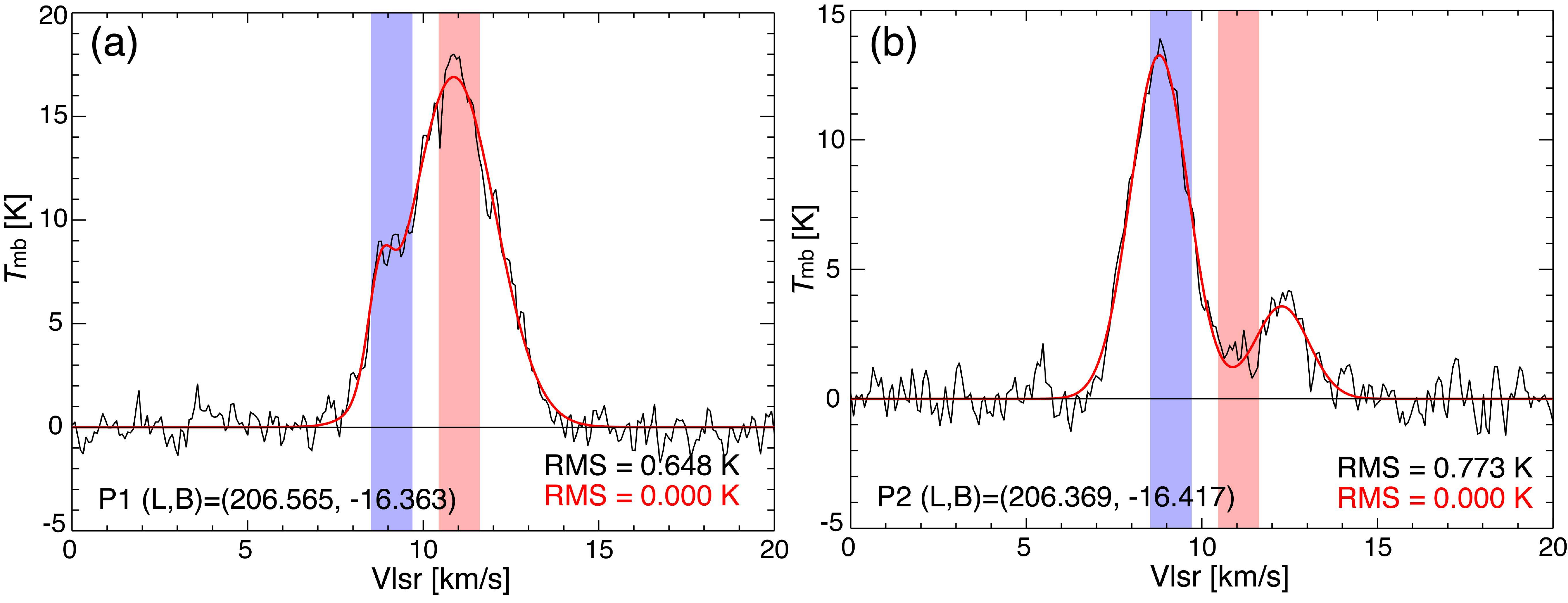} 
 \end{center}
\caption{\thirteencoh ~spectra toward the vicinity of P1 (a) and P2 (b) obtained with NANTEN2. The red line indicates the fitted profile by a double-gaussian function. The red and blue transparent belts represent the velocity ranges of the red and blue clouds, respectively.}
\label{gauss}
\end{figure*}
\begin{figure*}
 \begin{center}
  \includegraphics[width=17cm]{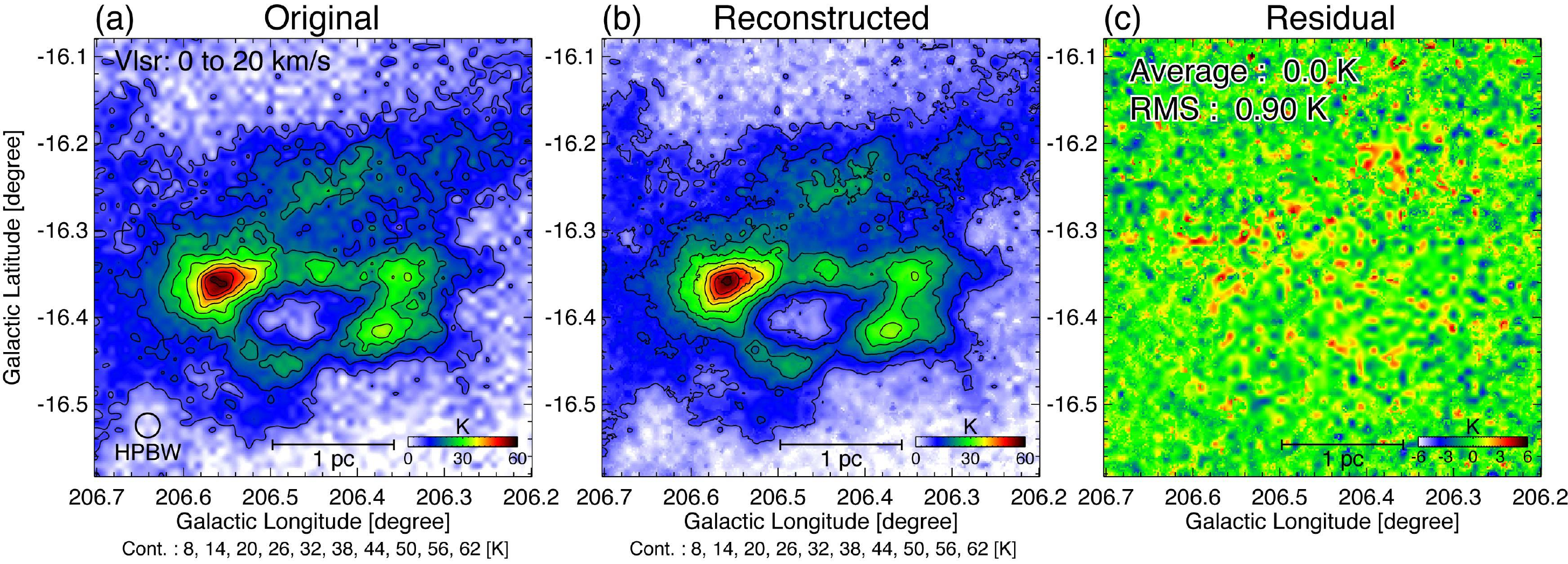} 
 \end{center}
\caption{(a) Integrated intensity distribution of \thirteencoh ~in the velocity range of 0 to 20 \kms. (b) Integrated intensity distribution of the reconstructed data obtained by fitting all line profiles in the data cube by double-gaussian functions. (c) Residual map of Figures~12a and 12b.}
\label{reconstruct}
\end{figure*}
Figure~11 shows that \thirteencoh ~line profiles in this region are well fitted by a double-gaussian function. We applied this fitting to all spectra and reconstructed a new data cube consisting of only the fitted double-gaussian profiles. In Figures~12a--c, we find the very good coincidence between the original data and the reconstructed data. The standard deviation of the residual map is $\sim$60 \% of that of the original data. We separated the double-gaussian line profiles of this reconstructed data into two single gaussian profiles of the lower velocity gaussian profile and the higher velocity gaussian profile. Here we call these two components blue cloud (estimated), and red cloud (estimated), respectively.

\begin{figure*}
 \begin{center}
  \includegraphics[width=11cm]{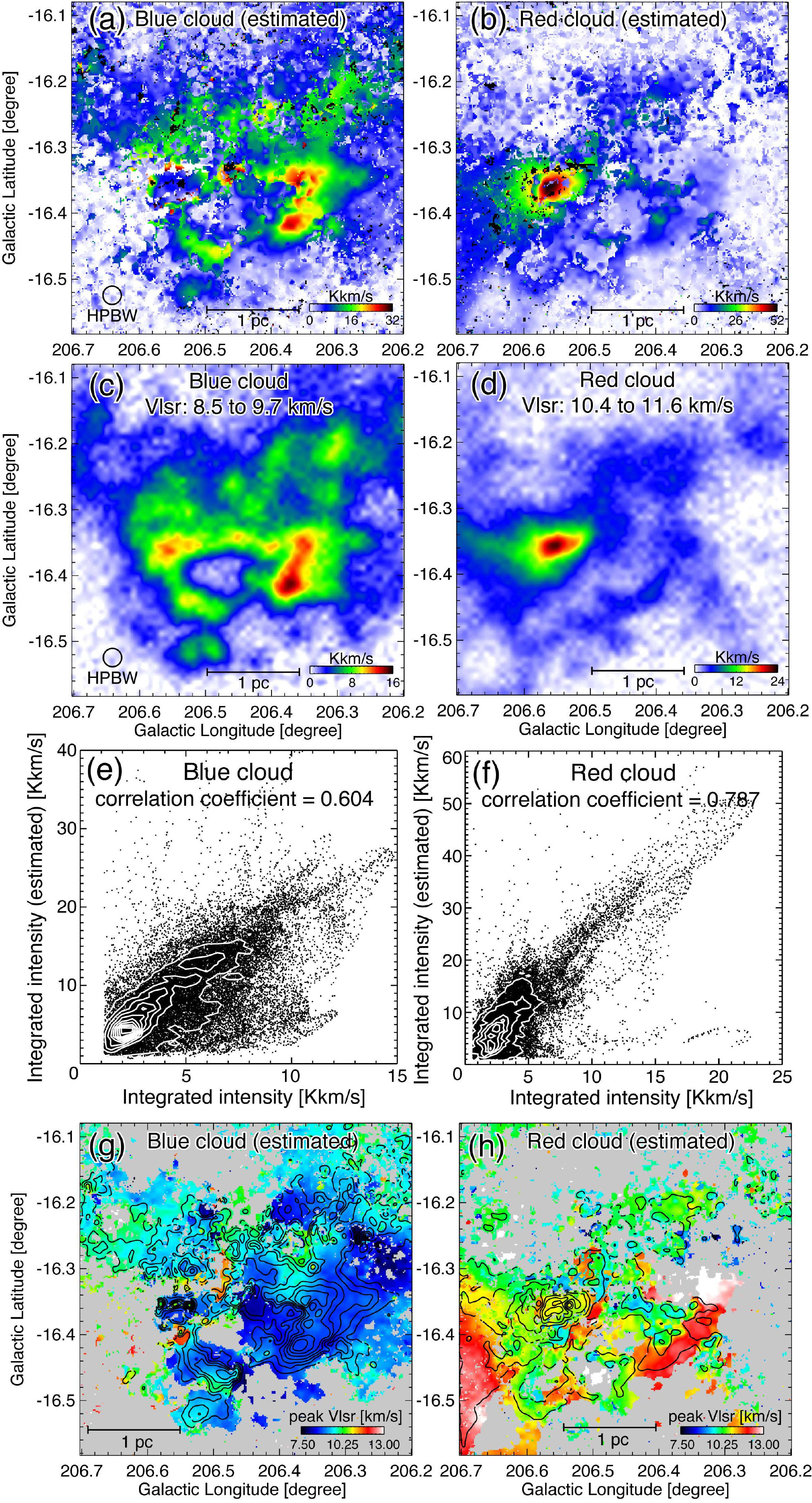} 
 \end{center}
\caption{(a, b) Integrated intensity distribution of the red cloud (estimated) and blue cloud (estimated), respectively. (c, d) Integrated intensity distributions of the red and blue clouds in \thirteencoh ~defined by the moments method. (e, f) Scatter plots between the distribution given by the moments method and the estimated distribution from the reconstructed cube for the blue and red clouds, respectively. The white contours indicate distribution of frequency of the number of pixels. (g, h) Peak velocity distributions of the blue and red clouds (estimated) overlaid with smoothed contours of integrated intensities obtained by figure~13a and 13b.}
\label{comparison}
\end{figure*}
The distributions of the blue cloud (estimated) and red cloud (estimated) are shown in Figures~13a and 13b. For comparison, we displayed distributions of the blue and red clouds in Figures~13c and 13d (same as Figures~5a and 5b), and took correlation between them (Figures~13e and 13f).
Pixels with $<$ 3 sigma were flagged in advance the correlation analysis. The correlation coefficients for the blue and red clouds are 0.604 and 0.787, respectively. Integrated intensities between the one obtained by the moments method and the estimated coincide within a factor of $\sim$2.
These results indicate that there are no significant morphological differences between the distribution of the estimated cloud and the one obtained by the moment method although the moments method may missed the half of the original integrated intensity. We therefore conclude that the moments method provides velocity ranges which represent clouds' major morphological features, and is useful also for the NGC~2024 region.
We show peak \vlsr ~distributions for the blue and red clouds (estimated) as Figures~13g and 13h for reference. The clouds' representative velocities derived by the moments method correspond blue to cyan (8.5 to 9.7 \kms) in Figure~13g and light green to red (10.4 to 11.6 \kms) in Figure~13h.




\end{document}